# Performance Verification of the FlashCam Prototype Camera for the Cherenkov Telescope Array

F. Werner*[1], C. Bauer[1], S. Bernhard[2], M. Capasso[3], S. Diebold[3], F. Eisenkolb[3], S. Eschbach[4], D. Florin[5], C. Föhr[1], S. Funk[4], A. Gadola[5], F. Garrecht[1], G. Hermann[1], I. Jung[4], O. Kalekin[4], C. Kalkuhl[3], J. Kasperek[6], T. Kihm[1], R. Lahmann[4], A. Marszalek[7], M. Pfeifer[4], G. Principe[4], G. Pühlhofer[3], S. Pürckhauer[1], P.J. Rajda[7], O. Reimer[2], A. Santangelo[3], T. Schanz[3], T. Schwab[1], S. Steiner[5], U. Straumann[5], C. Tenzer[3], A. Vollhardt[5], D. Wolf[5], and K. Zietara[7] for the CTA consortium[8]

[1]*Max-Planck-Institut für Kernphysik, PO Box 103980, 69029 Heidelberg, Germany*
[2]*Institut für Astro- und Teilchenphysik, Leopold Franzens Universität Innsbruck, Technikerstrasse 25/8, A 6020 Innsbruck, Austria*
[3]*Institut für Astronomie und Astrophysik, Abteilung Hochenergieastrophysik, Kepler Center for Astro and Particle Physics, Eberhard Karls Universität, Sand 1, D 72076 Tübingen, Germany*
[4]*Physikalisches Institut, Friedrich-Alexander Universität Erlangen-Nürnberg, Erwin-Rommel-Str. 1, D 91058 Erlangen, Germany*
[5]*Physik-Institut, Universität Zürich, Winterthurerstrasse 190, 8057 Zürich, Switzerland*
[6]*AGH University of Science and Technology, Al. Mickiewicza 30, 30-059 Krakow, Poland*
[7]*Astronomical Observatory, Jagiellonian University, ul. Orla 171, 30-244 Krakow, Poland*
[8]*Full consortium author list at* https://www.cta-observatory.org.

**Abstract**

The Cherenkov Telescope Array (CTA) is a future gamma-ray observatory that is planned to significantly improve upon the sensitivity and precision of the current generation of Cherenkov telescopes. The observatory will consist of several dozens of telescopes with different sizes and equipped with different types of cameras. Of these, the FlashCam camera system is the first to implement a fully digital signal processing chain which allows for a traceable, configurable trigger scheme and flexible signal reconstruction. As of autumn 2016, a prototype FlashCam camera for the medium-sized telescopes of CTA nears completion. First results of the ongoing system tests demonstrate that the signal chain and the readout system surpass CTA requirements. The stability of the system is shown using long-term temperature cycling.

## 1. Introduction

CTA is an international effort to build the next-generation ground-based gamma-ray observatory [1]. The observatory will consist of two arrays with several dozens of telescopes each to cover the photon energy range from tens of GeV to hundreds of TeV on both hemispheres [2].[1] The FlashCam group has developed a camera concept for CTA telescopes based on Ethernet readout of purely digitally processed photosensor signals and constructed several evaluation setups over the last years to verify it [4, 5].

In autumn 2016, a full-scale prototype of a photomultiplier tube-based FlashCam camera for the medium-sized telescopes (MST, see Ref. [6]) of CTA will be completed. The main aim of the prototype is to verify all safety, operational and performance aspects of the system before the pre-production and eventual mass production of cameras. In the following, the design of FlashCam is introduced, and the measurement methods and first results of the test setup used to verify the physics performance under realistic environmental conditions are described.

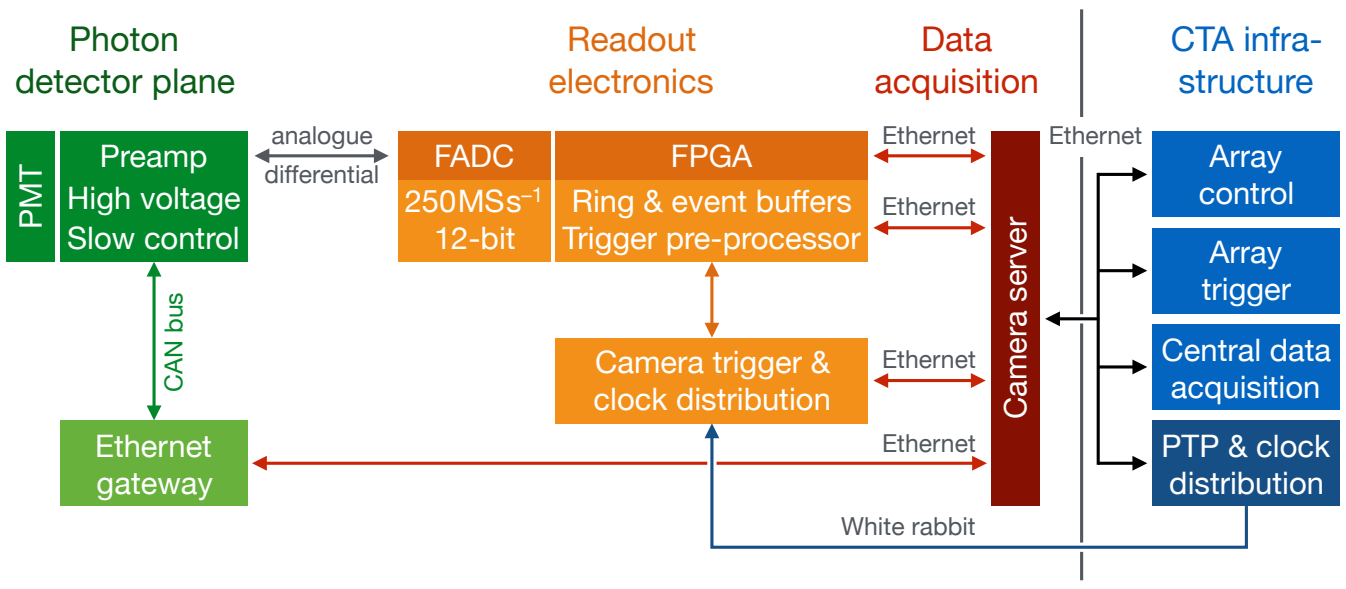

**Fig. 1.** Basic building blocks of the FlashCam signal chain (left) and interfaces to the central CTA infrastructure (right). See text for a detailed description.

## 2. FlashCam design

The design of FlashCam follows a horizontal architecture, with the photon detector plane (PDP), the readout electronics (ROS),

*Corresponding author. *E-mail address:* Felix.Werner@mpi-hd.mpg.de.

[1]The concept and goals of CTA are introduced in the context of multi-messenger astronomy in contribution [3] to this issue.

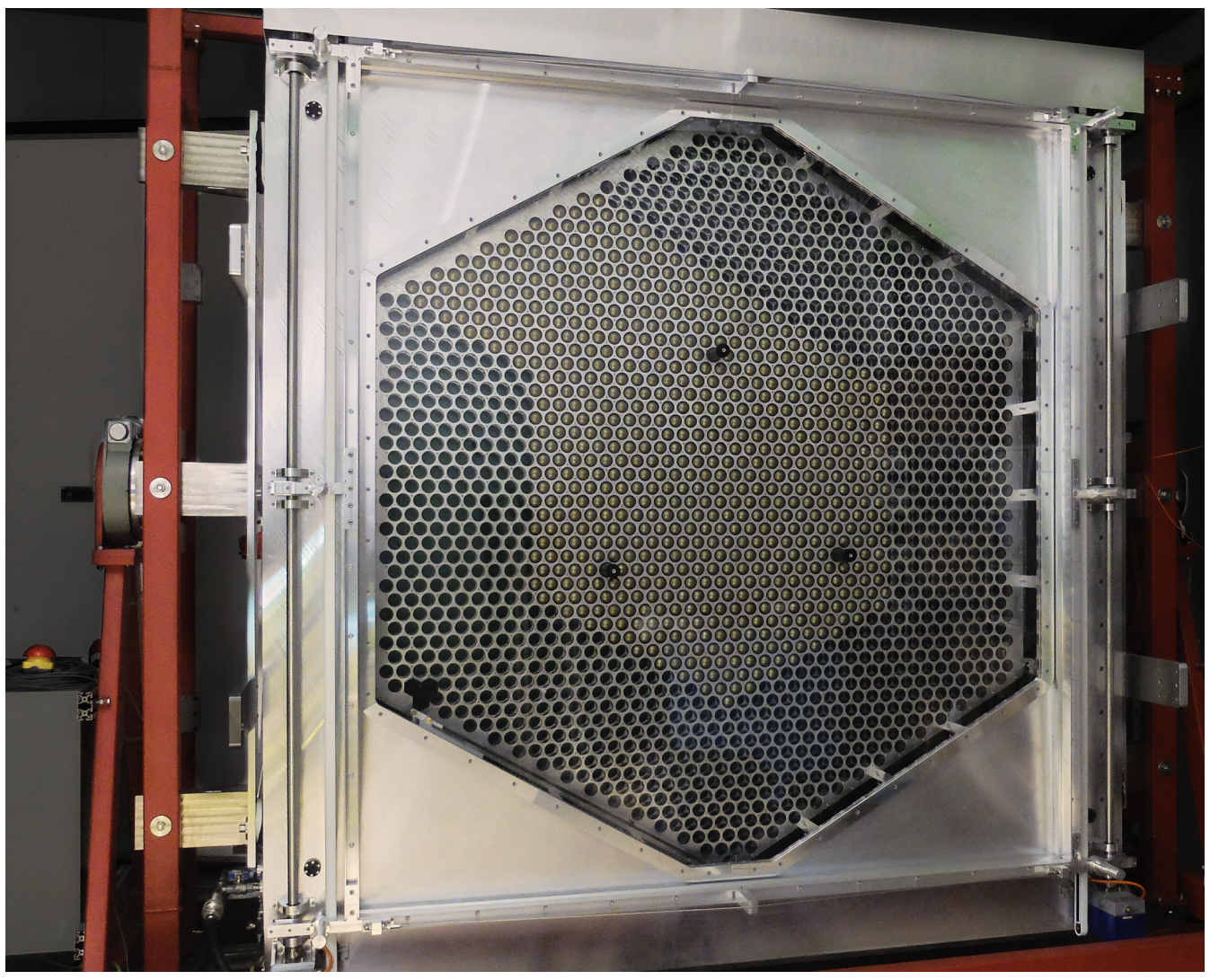

**Fig. 2.** FlashCam prototype camera for MST with opened shutter. Part of the hexagonal photon detector plane is equipped with PMTS, the rest with dummy heating modules. The camera is mounted in a stable rotatable frame with attached programmable industrial servo (left) for mechanical stress tests. The cooling circuit and dry air supply are attached to the interface panel on the lower left, power and Ethernet fibres on the lower right.

and the data acquisition system (DAQ) as key building blocks, see Fig. 1. The PDP contains photomultiplier tubes (PMTS) arranged in a hexagonal structure with 50 mm pixel spacing and is composed of 147 modules with 12 PMTS each. Each PDP module contains a DC–DC converter to provide high voltage, pre-amplifiers as well as a CAN bus interface for slow control, monitoring, and safety functions. A non-linear amplification scheme is used, in which signal amplitudes >250 photoelectrons (p.e.) saturate in a controlled way (with the integral growing logarithmically with input charge) to extend the dynamic range up to >3000 p.e. while retaining linearity and sub-p.e. resolution for small signals. The analogue signals are transmitted differentially via cat. 6 shielded twisted-pair cables to the readout electronics, the design of which is based on a fully digital approach with continuous signal digitisation. The signals are sampled with 12-bit FADCS at a rate of 250 MS $s^{-1}$. The samples are buffered in FPGAS and processed in a configurable way to derive a trigger decision which is typically optimised for localised, short light pulses. Waveforms comprising an adjustable time slice (up to 15.6 µs and with a configurable offset relative to the time of trigger) of these camera-level triggers are read out via a camera-internal, high performance Ethernet network, using off-the-shelf switches and a standard commercial server. This camera server is typically located at a central computing cluster and connected directly to the camera via up to four 10 Gbit Ethernet fibres. Custom-developed software implements the high performance front-end to back-end data transfer, event building, optional zero-suppression, event selection, exchange of array trigger information, and data formatting for the array-wide data acquisition. Two additional Gbit Ethernet fibres to the camera are used for slow control and monitoring (direct fibre between camera server and camera), and an array-wide precision clock distribution network (private White Rabbit network). The camera server further provides interfaces to the CTA-wide array control, data acquisition, and to the software-based array trigger.

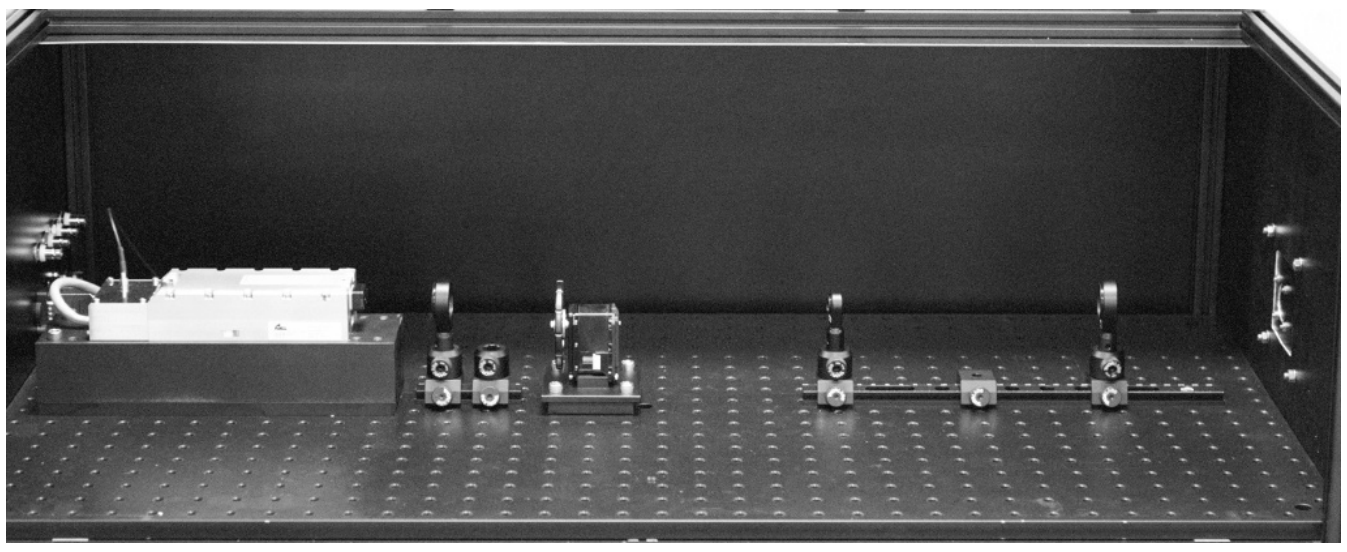

**Fig. 3.** Light source used for testing the prototype camera. Left to right: pulsed 355 nm laser, OD1.3 filter, motorised OD4 filter wheel, 10× beam expander, diffuser system. Not shown: UV LED to produce a programmable DC illumination level and baffles to block reflections within the enclosure.

# 3. Experimental setup

The prototype camera has been equipped with the complete readout electronics for 1764 channels along with 60 PDP modules (totalling 717 pixels) from a first prototype batch, see Fig. 2. PDP modules with two different PMT variants under evaluation, Hamamatsu R12992-100 with 7 dynodes and Hamamatsu R11920-100 with 8 dynodes,[2] are distributed evenly over the detector plane. The readout electronics is connected to a server via two commercial Ethernet switches in the camera and four bundled, 1 km long 10 Gbit single-mode Ethernet fibres. A set of control and monitoring processes further communicates with the safety and slow control systems via a separate Gbit fibre.

During operation, the interior temperature of the camera is stabilised using water-air heat exchangers connected to an external, programmable thermocirculator. Humidity inside the camera is kept low using a constant inflow of dry air to avoid condensation, particularly at low temperatures. Similar facilities are foreseen for on-site operation.

A light source consisting of a pulsed 355 nm laser, a programmable OD4 filter wheel, and a DC UV LED, see Fig. 3, has been used to evaluate the camera at realistic pulsed intensities (0.3–10,000 p.e. per pixel) and DC illumination levels (0–3 GHz p.e. per pixel). The electrical time synchronisation signal of the laser unit is digitised in the camera readout electronics for triggering. Its shape is further used to derive the event-wise, random phase delay between the sub-nanosecond light pulses and the sampling clock with an RMS of <50 ps to separate the phase jitter from the intrinsic time jitter of the signal chain. Before any measurement run, the camera undergoes an automatic flat-fielding procedure to equalise the end-to-end gain across the photon detector plane with an RMS of <3%.[3] Calibration parameters are obtained from maximum

[2] The two variants have been compared extensively and either provides adequate performance for FlashCam. In concordance with the preferences of other camera systems within CTA, the FlashCam pre-production cameras will be equipped with 7-dynode PMTS.

[3] The reconstructed average charges show an RMS of 4% across the detector plane (this includes spatial inhomogeneity of the illumination, and the variation of quantum and collection efficiencies of the PMTS). Thus, flat-fielding of all pixels is done at a single fixed intensity within the linear

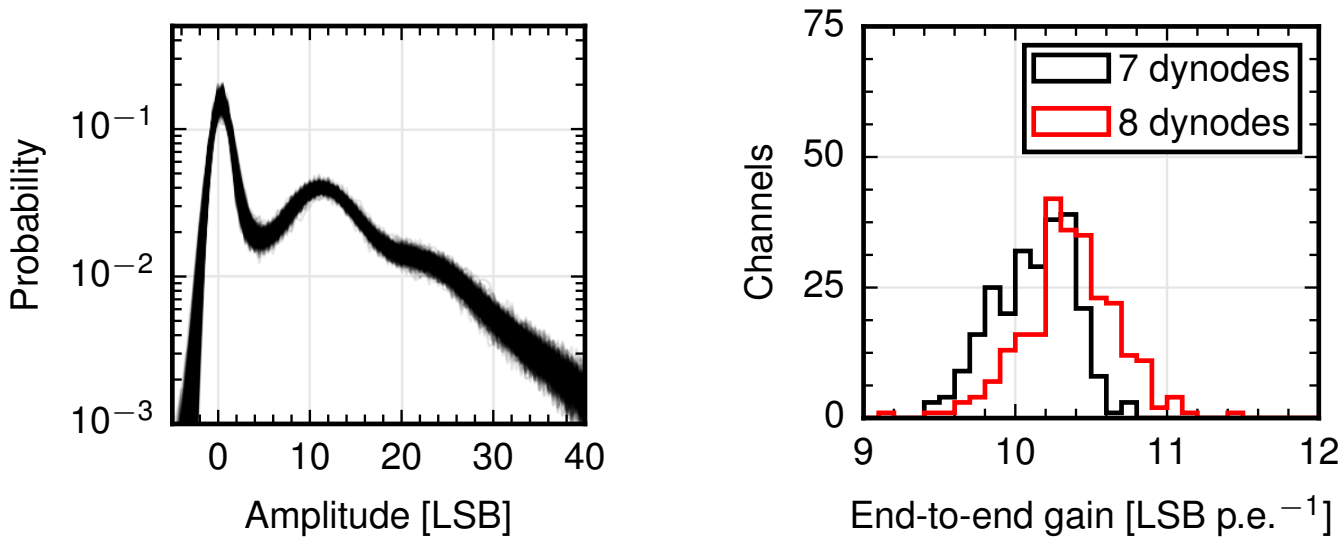

**(a)** SPE spectra obtained at an average charge of 0.9 p.e. per pixel and pulse, overlaid for 510 channels.

**(b)** End-to-end gain parameters from maximum likelihood fits to the SPE spectra.

**Fig. 4.** Single-photoelectron (SPE) spectra at nominal gain **(a)** and extracted end-to-end gain in the linear region **(b)** of 510 channels.

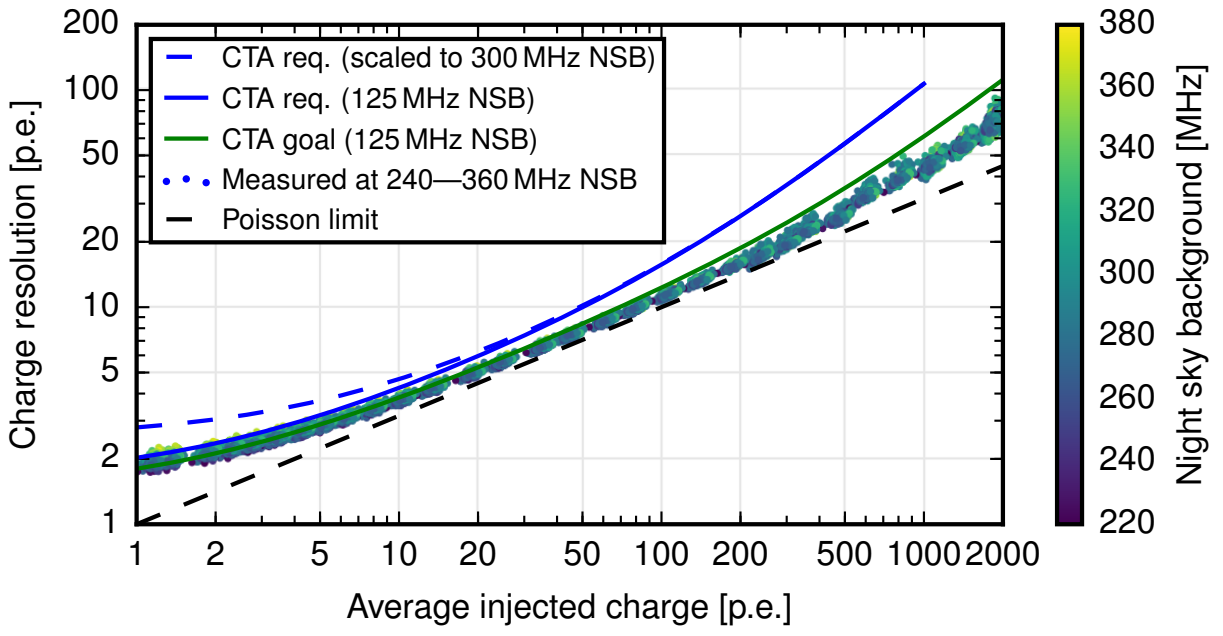

**Fig. 5.** Measured charge resolution of 450 individual channels at dark sky DC illumination level as a function of average injected charge (pulsed laser). The CTA requirements, originally formulated for a somewhat low DC illumination of 125 MHz p.e. per pixel (upper solid curve), are met for all channels.

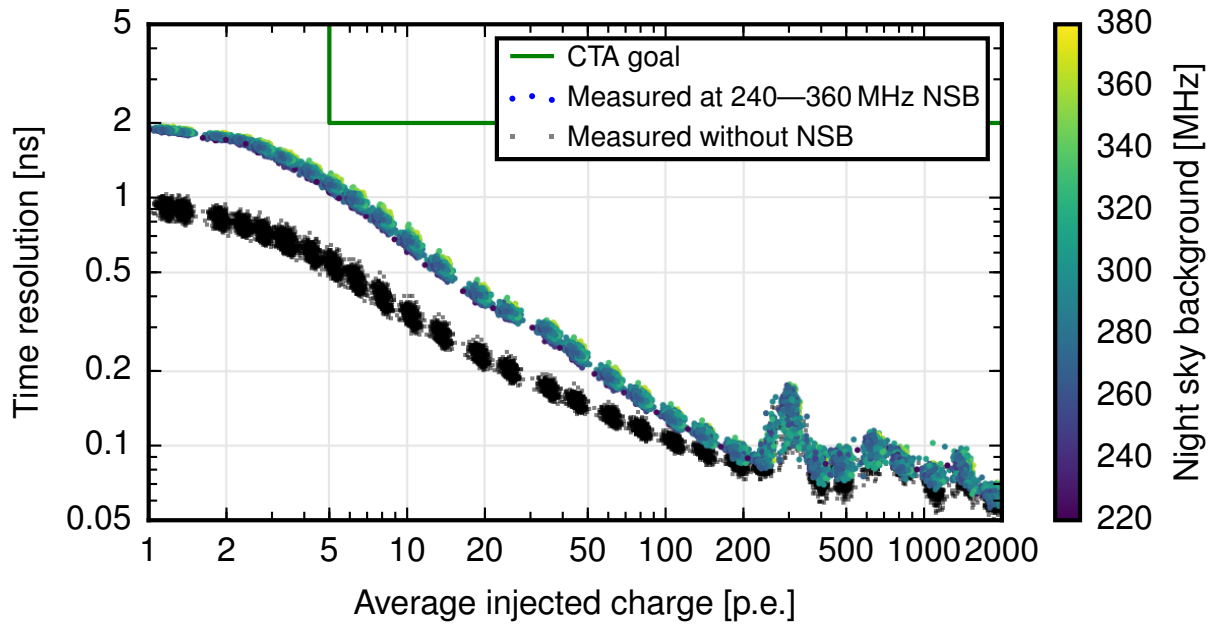

**Fig. 6.** Measured time resolution of 450 individual channels at zero (lower band) and dark sky (upper band) DC illumination level as a function of injected charge (±2 ns search window). At low amplitudes, the resolution is dominated by statistical noise and the transit time spread of the PMTs. Around 250 p.e., the resolution is dominated by the transition from linear to non-linear amplification. The CTA goal resolution of 2 ns above 5 p.e. is met for all channels.

likelihood fits to single-photoelectron (SPE) spectra, see Fig. 4, and expanded to the non-linear regime of the signal chain (>250 p.e.) using non-parametric models of the calibration curve. The systematic uncertainty of the resulting charge reconstruction is expected to be <10%.[4]

## 4. Verification measurements

The CTA observatory follows a verification-based product acceptance procedure: all products to be deployed on site must fulfil a list of environmental, RAMS[5], and performance requirements. In the following, the first results of system-level verifications of the performance and stability are shown, introducing the requirements before discussing the test results.

**Readout system.** The readout system of an MST must be able to capture events arriving at random with a rate of 4.5 kHz with <5% dead time. Further, a goal readout rate of 9 kHz is formulated since triggering deeper in the noise will reduce the energy threshold of the MST subsystem and short-term bursts, e.g., due to upward fluctuations in the night sky background (NSB), should not lead to excessive dead time affecting shower statistics and stereo trigger. The fully equipped readout system of the FlashCam prototype (1764 FADC channels and 96 trigger channels) can be read out with a bulk data rate of >3.3 GByte s$^{-1}$ via the installed 1 km fibre bundle, at about 85% usage of a single 2012 CPU core (Xeon E5-2690) including event building. For the target waveform duration of 88 ns, readout rates of >35 kHz (random rate) can be reached without inducing dead time. Hence, outperforming the required readout rate by >7× with no dead time, the FlashCam readout system is well suited for reaching a consistent readout performance under varying conditions and, e.g., observations with low energy threshold.

**Performance of the signal chain.** The fundamental performance parameters of the signal chain are the quality of the pulse reconstruction and its dynamic range. The camera is required to perform charge measurements up to 1000 p.e. with a given resolution. Furthermore, an extended dynamic range up to 2000 p.e. is desirable for high-energy events along with a per-pixel RMS time resolution of 2 ns for signals >5 p.e. Note that the camera is subject to a significant statistical background during observation: in dark sky conditions, a background rate of about 300 MHz p.e. per pixel is expected, while brighter sky regions and celestial objects can increase the background by a factor of five or more. Therefore, the charge resolution and time resolution of the FlashCam prototype have been evaluated at NSB rates up to 3 GHz – Figs. 5 and 6 illustrate the performance at dark sky NSB rates. The signal reconstruction outperforms all requirements and goals due to careful design of the front-end, low-noise digitisation, and matched digital signal processing on the camera server.

**Long-term and temperature stability.** In-field, the camera must be functional for observation when subject to ambient temperatures of −15°C to 25°C. Although the interior temperature of the camera will be stabilised to a certain degree by an external thermocirculator, the camera – and its signal chain in particular – should be robust against temperature changes as this will lead to higher quality data and less down time. Also, relaxed requirements on the thermocirculator will likely lead to lower investment and maintenance costs for the observatory.

The stability of the signal chain over time and temperature change was assessed by cycling the temperature of the coolant over a range of 5°C to 35°C in hourly steps of 2 K over a period of 30 h, with resulting interior temperatures of about 13°C to 41°C. When the interior temperature reached thermal region, typically with 10–50 p.e. pulses and using the relative fluctuations of the measured signals as a measure of gain.

[4] Work towards an article describing the digital signal processing, reconstruction, and calibration in detail is ongoing.

[5] Reliability, Availability, Maintainability, and Safety.

equilibrium after each step, light flashes of several fixed intensities between 1 p.e. and 300 p.e. were measured, focusing mainly on the linear regime of the signal chain. The camera system itself and the core reconstruction parameters were found to be well-behaved over the entire range of operation and no significant hysteresis was observed. In Fig. 7, the most significant temperature dependencies of the signal chain are shown, namely for baseline, timing, and gain. The baseline is used for an accurate, per-event measurement of the DC illumination level of each pixel. At nominal gain, its average temperature coefficient is equivalent to about 2 MHz p.e. $K^{-1}$, less than 1% $K^{-1}$ at typical NSB rates, and can be corrected during reconstruction using a lookup table. The main source of temperature dependencies in timing and gain can be explained by a temperature drift of the HV applied to the PMTs and leads to average temperature coefficients of about 3 ps $K^{-1}$ in timing and −0.2% $K^{-1}$ in gain[6] for the preferred PMT variant. These effects may be corrected either by adjusting the HV based on temperature measurements at the PDP or during reconstruction using a lookup table. No striking effects were observed in the non-linear transition region, but more measurements at signal amplitudes >250 p.e. are required to evaluate the stability of the full calibration curves.

## 5. Conclusions and outlook

The FlashCam concept is currently being implemented for the medium-sized telescopes of CTA, with a full-scale prototype camera nearing completion in autumn 2016. The first system-level performance verifications have shown that both concept and implementation are excellent options for modern Cherenkov cameras. The readout system, based on FPGAs, commercially available Ethernet components, and a high-performance camera server, has been shown to capture events at rates >35 kHz without inducing dead time. This leaves a comfortable margin for fluctuating backgrounds and low-threshold observations. Due to the careful design of the signal chain, the quality of the pulse reconstruction outperforms all requirements and goals of the CTA over the complete dynamic range up to 2000 p.e. and beyond. First long-term temperature cycles of the full camera have shown that all components work properly over an extended range of operation. Furthermore, the temperature dependencies of the signal path are small, deterministic, and thus easily accounted for.

The next major steps in this work are the verification of the trigger performance and further long-term testing of the complete system under realistic conditions. In parallel, work is ongoing to implement and verify the software interfaces to the array and towards the pre-production of two cameras in 2017.

## Acknowledgments

We gratefully acknowledge support from the agencies and organisations found under Funding Sources at `https://www.cta-observatory.org`.

[6] The gain change found with SPE fits is consistent with the change of the relative fluctuations of the reconstructed charge at different illumination levels.

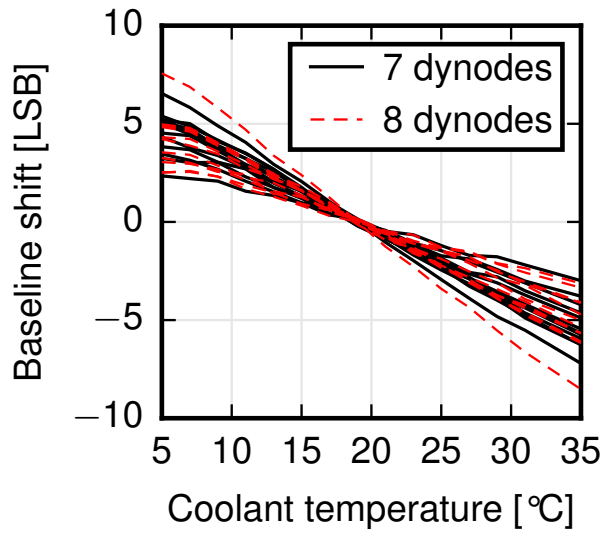

**(a)** Change of measured DC level as a function of coolant temperature.

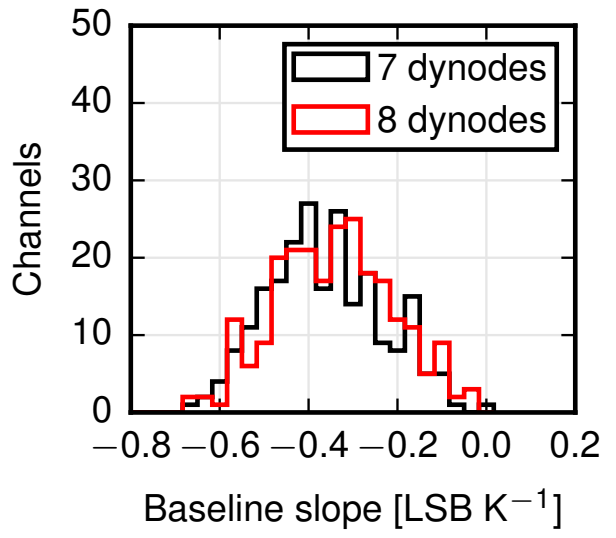

**(b)** Linear temperature coefficients of the measured DC level.

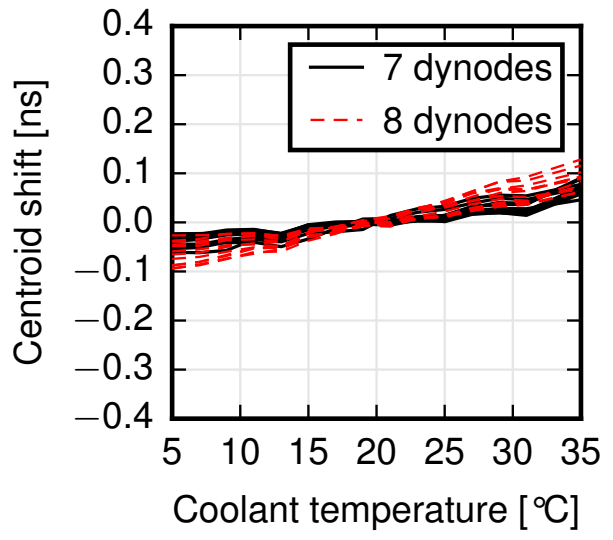

**(c)** Change of measured arrival time of light pulses as a function of coolant temperature.

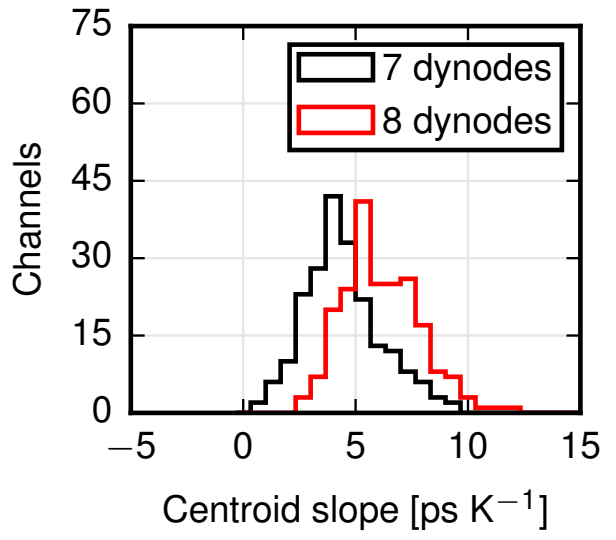

**(d)** Linear temperature coefficients of the measured arrival time of light pulses.

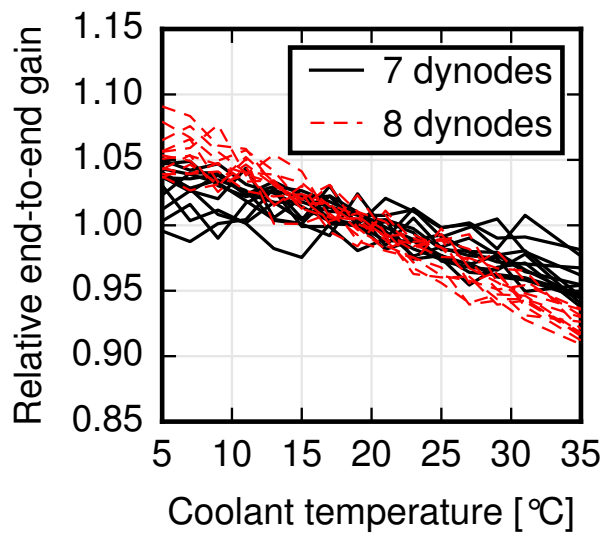

**(e)** Change of end-to-end gain as a function of coolant temperature.

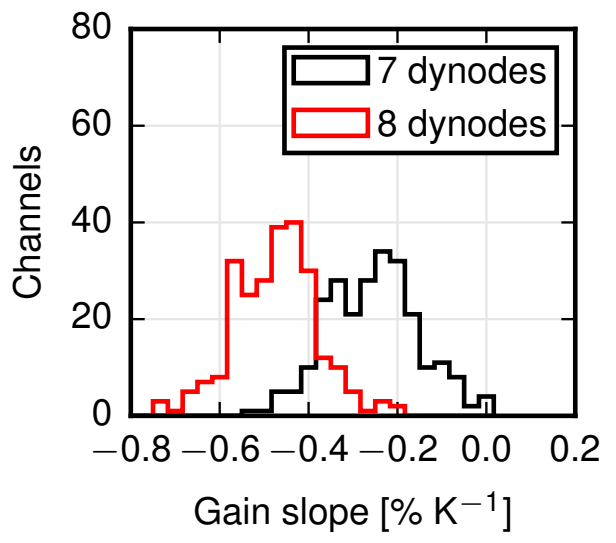

**(f)** Linear temperature coefficients of the end-to-end gain.

**Fig. 7.** Temperature dependencies of the reconstruction parameters baseline (a–b), timing (c–d), and gain (e–f, from SPE spectra). For display purposes, the left-hand figures show the temperature dependencies only for a few channels of each PMT variant. The right-hand figures show histograms for all channels.